# A Consideration of Cosmic Evolution from the Points of View of the Inflationary and Cyclic Theories

Anantya Bhatnagar

This study reviews the advances made in inflationary theory, especially regarding the seeming disparity between inflation energy and dark energy, and their significance to cosmic evolution as a whole. I attempt to connect the two sources of expansion and thereby enhance the predictive capacity of the consensus model. I also attempt to contrast its strengths and weaknesses with those of the cyclic theory developed by Turok and Steinhardt, particularly where cosmic evolution over larger time scales is concerned. Furthermore, I endeavor to provide a physical meaning to the existence and workings of dark energy as well as dark matter in the cyclic model (as compared to their simply being extra parameters in the inflation picture), so that each observed component of the universe is given significance.

## I. Introduction

The widely accepted cosmological picture today has taken many varied turns in order to have become so established. From Einstein's early classical relativistic model being assumed to imply a big bang through Hawking and Penrose's work with the singularity theorems to the inflationary model that attempts to explain the homogeneity and isotropy that characterizes the universe through a brief and rapid epoch of inflation, myriad advances in the paradigm have occurred. Even cyclic models were proposed early on that worked on a closed universe – based on spatial curvature – but these were discarded due to incompatibilities with the measured matter density of the universe (in relation to its critical density) and hence its scale factor. However, one of the biggest problems that still remain in our understanding of the cosmos is the physical nature of the invisible energy that seems to pervade over 70% of the universe and appears to be the reason for its current accelerating expansion.

One of the problems with the inflationary model that particularly stands out is that it did not seem to predict dark matter or, in particular, dark energy as part of the theory. Rather, these additions were made to the theory when discovered through experimental measurements, which in my view makes it quite cumbersome. In this work I shall attempt to evaluate the mechanism by which cosmic evolution is theorized to occur, using inflation and dark energy in the inflationary model, and compare this against the result obtained from a similar evaluation with an alternative viewpoint. This will be based on the cyclic model developed by Turok and Steinhardt, which shall be put forth qualitatively as a physically intuitive picture of the universe.

## II. A review of the role of dark energy in the overall picture as it stands today

Dark energy, as hypothesized in the present, was absent from the original inflationary picture and, not being a requirement for the theory to work, was added to the model later on, once it was clear that the astronomical observations and experimental results did not match with the predictions being made by the theory as it stood for the current stage of the universe's evolution. Nevertheless, once it was established that dark energy had to be incorporated into the framework, it was theorized to take one of two possible forms. The first is found in Einstein's original field equations –

$$R_{\mu\nu} - \tfrac{1}{2}g_{\mu\nu}R + \Lambda g_{\mu\nu} = 8\pi G T_{\mu\nu} \qquad - \quad (1)$$

Here, Λ is theorized to take some form of repulsive quantity that is constant over spacetime, since it is coupled to the metric itself. However, the first problem that this poses as a candidate for dark energy is that it is invariant as the universe evolves from one stage to the next, so it is difficult to imagine how it would be able to explain the varying rates at which the universe has been expanding in the radiation-, matter- and quintessence-dominated periods. Some progress has been made in finding out the dynamics of a time-varying constant $\Lambda \to \Lambda(t)$ – see for example [1] – and there is hope that more can be understood in the future about the possible role of Λ as a form of dark energy.

This brings us to the second form that dark energy is currently theorized to take. The quintessence is thought to be an all-pervading ether-like entity that, at this stage of the universe's evolution, is dominating over matter and radiation, and so is able to cause an accelerated cosmic expansion. It is expressed as a scalar field with the following general equation of state [2] –

$$w_q = \frac{p_q}{\rho_q} = \frac{\frac{1}{2}\dot{Q}^2 - V(Q)}{\frac{1}{2}\dot{Q}^2 + V(Q)} \qquad - \quad (2)$$

While this quantity is simply added to the inflationary model due to the latter's flexibility in dealing with a large number of parameters (and it must be noted that the theory works just as well without the presence of this experimentally deduced addition), a rather different picture is painted in the cyclic model [3], on which the analysis in this paper is based.

Contrary to the previous oscillating cosmic theories which depended on spatial curvature and therefore involved a closed universe which transitioned from big bang to big crunch, the cyclic model proposed by Turok and Steinhardt works on a flat universe at the point of transition to the next cycle. While previous theories interpreted the cycle as an infinite number of singularities occurring periodically and causing the re-genesis of the universe from nothingness, the shift from big crunch to big bang here is taken as the temporary collapse of an extra dimension due to the collision and rebounding of two branes, therefore causing the interbrane separation to go to zero and increase again, generating matter and radiation at the point of the bounce. In contrast to the role that it plays in inflation, dark energy is absolutely essential to the cyclic picture, in causing the accelerated cosmic expansion that brings the universe's matter and radiation density down, so that the interbrane separation is allowed to reduce by way of the 'fifth force' – the interaction between the two branes. Hence, mathematically, dark energy is hypothesized to take the form of a potential $V(\phi)$, where $\phi$ is a scalar field that determines the interbrane distance [3]. According to the analysis of Turok and Steinhardt in [3], this potential can be taken as

$$V(\phi) = V_0(1 - e^{-c\phi})F(\phi) \qquad - (3)$$

Current observational limits place the value of c at 10.

Physically, the stage at which this potential is actually effective is the current period of acceleration in the universe's expansion, so this dark energy represents the quintessence-dominated stage. This may be interpreted as a permeating ether-like quantity in the higher dimensional metric. This leads into the time immediately preceding the bang, when this potential appears to play a non-trivial role. Here, the two branes are stretched out, and made flat and smooth (and this is the point that sets this cyclic model apart from the currently accepted inflationary picture, since the energy scales involved in the mechanism that allows for the homogeneous, isotropic universe observed today, are far lower than those in inflation respectively). In the following sections, it shall be demonstrated that this form of cyclic theory can be a physically intuitive picture of the universe, and so it follows that, as dark energy plays such a key role in the theory, it should likewise be presented as an elegant ingredient in the metric.

## III. An analysis of the inflationary picture, especially from a physical point of view

The standard cosmological model agrees that there was a point in time approximately 13.8 billion years ago, when the universe was of infinitely high density which, in the thermal sense of entropy, corresponds to a low entropy state. A period of rapid inflation is then predicted, allowing the universe to become extremely homogeneous due to the speed of expansion of the spacetime. After this epoch, the formation of large scale structures due to gravitational clumping and clustering seems to reduce the entropy of the system, as structures become more ordered. However, when this situation is looked at closely, it is seen that an expanding, self-gravitating gas increases in potential energy and so reduces in kinetic energy (through conservation of total energy). This reduces its temperature and hence the entropy. Conversely, a contracting gas of similar nature will increase in entropy, thereby allowing for large-scale, well-defined structures to form.

The point in the current picture where inflation is relevant, the very early universe, does not provide us with a big problem in terms of entropy evolution. After the inflaton field is created and it plays its part in causing the universe to expand extremely rapidly, it decays and the universe reheats. However, perturbative reheating, as suggested by Turner, Steinhardt, Albrecht and Wilzcek [4], is slow and produces a low reheating temperature, suggesting that the increase in entropy per unit time could be lower than expected after such a brief and rapid expansion phase. However, this is not much of a problem because fluctuations are absent in this inflaton field's equation of motion, a point that is resolved in [5] and [6].

In this picture, however, the second point of interest is the current epoch of cosmic acceleration that is being dominated by the pervasive dark energy. While this is clearly increasing the entropy of the universe as a system, it has been known for some time that the universe is expanding, which in any case leads to an increase in entropy. In this form, therefore, dark energy is another parameter that has been inserted to comply with experimental observations.

There is a barrier that the introduction of this parameter provides, while viewing the inflationary picture as a whole, over a large time frame – there seem to be two forms of repulsive energy that are not connected in any way. The first is the inflaton field, which drives the initial period of extremely rapid expansion, and the second is dark energy, which is fueling the current slower, yet accelerated expansion of the universe. If we want to view these two phases as being somehow connected, we need to attempt to visualize and see how one phase can logically lead into the other. Otherwise, only two other scenarios are probable here – one, where the dark energy field was created at the initial moment of singularity, in which case the rate of inflation should have been more than what it is theorized to be, and the other, where it existed before the singularity. In inflationary theory, however, the second scenario becomes meaningless in itself.

We are left with two situations to consider. The first is the dark energy/quintessence field being created along with matter and radiation in the reheating phase of the initial inflationary period. However, this gives rise to a natural question – if it were created at that point, why would it not affect cosmic expansion immediately? After reasoning with a qualitative perspective, it appears to me that it is perhaps because the process of the quintessence field causing acceleration is only effective at larger spatial scales, or is only activated once radiation and matter domination have reduced. However, this would again require fine-tuning of the theory, which reduces its physical impact. Nevertheless, a counter-argument to this can also be examined, namely, what makes the creation of such a field, which is inactive until the present time, possible? While fine-tuning could be required, it is also possible that this could already be present as fluctuations, meaning that this phenomenon could be physically manifest in the theory without the need for any manual adjustments. It is quite improbable though, that these fluctuations could be manifest somehow without actually being inserted by hand (and fine-tuning the values), and so this possibility is more or less discarded.

The second possible mechanism is that the dark energy field that dominates proceedings today could be the same inflaton field but at a lower energy scale, due to loss of energy through generation of matter and radiation. This is a rather more attractive idea as it would allow the inflationary model to be a more interconnected, and hence more elegant, picture. Also, the entropy changes would be far less drastic than those involved in the creation of a separate field, making this more physically reasonable than the first possibility. A fundamental analysis of the dynamics of the inflaton field will hopefully help reveal whether this scenario is probable or not. Classically, the equation of motion for the inflaton field is known to be [7] –

$$\ddot{\phi} + 3H\dot{\phi} - \nabla^2\phi + \frac{dV}{d\phi} = 0 \qquad - (4)$$

I shall proceed by attempting to make a comparison with (2) which, as previously, described, is the equation of state for the dark energy field – both possibilities will be discussed. Firstly, for the cosmological constant $\Lambda$,

$$w_q = \frac{p_q}{\rho_q} = \frac{\frac{1}{2}\dot{Q}^2 - V(Q)}{\frac{1}{2}\dot{Q}^2 + V(Q)} = -1$$

$\dot{Q} = 0 \rightarrow Q(t) = const.$, as expected

For a quintessence field though, Q is dynamic and so $w_q$ varies with time. The extreme case, where $w_q < -1$, is known to hypothetically end in a Big Rip, but this scenario is incompatible with current measurements of the scale factor of the universe. On the other hand, it is also known that any equation of state $w_q < -\frac{1}{3}$ will cause an accelerated expansion of the universe, and so limits are placed on this parameter, specifically

$$-1 \leq w_q < -\frac{1}{3} \qquad -(5)$$

Adding the previous expression for $w_q$ into (5),

$$-1 \leq \frac{\frac{1}{2}\dot{Q}^2 - V(Q)}{\frac{1}{2}\dot{Q}^2 + V(Q)} < -\frac{1}{3} \qquad -(6)$$

If the two inequalities are taken separately,

$$\frac{\frac{1}{2}\dot{Q}^2 - V(Q)}{\frac{1}{2}\dot{Q}^2 + V(Q)} \geq -1 \rightarrow \dot{Q}^2 \geq 0 \qquad -(7)$$

$$\frac{\frac{1}{2}\dot{Q}^2 - V(Q)}{\frac{1}{2}\dot{Q}^2 + V(Q)} < -\frac{1}{3} \rightarrow 2\dot{Q}^2 - 2V(Q) < 0 \rightarrow \dot{Q}^2 < V(Q)$$

$$0 \leq \dot{Q}^2 < V(Q) \qquad -(8)$$

Again, taking the second inequality above,

$$\frac{d}{dQ}(\dot{Q}^2) < \frac{dV}{dQ}$$

$$2\left(\frac{d}{dQ}(\dot{Q})\right)(\dot{Q}) < \frac{dV}{dQ}$$

$$\frac{dV}{dQ} > 0 \qquad -(9)$$

A comparison of (8) and (9) with (4) shows that the bounds put on the quantities in the dark energy field's equation of state are more or less compatible with the inflaton field's equation of motion. This can be accompanied by the reasonable explanation that despite the loss of energy from the inflaton field in the reheating phase of the initial inflationary period, $V(Q)$ and $\dot{Q}$ remain within an expected range. Therefore, quantitatively speaking, it could be possible that the inflaton field itself could lose energy, but not decay entirely, and be present as the dark energy that is driving the accelerated expansion observed today.

There is another complication that dark energy creates in this period. As it works in inflationary theory, it allows for a wide range of evolution states in the long run, the most probable amongst which is a completely vacuous universe that will exist as an empty space for eternity. Nevertheless, cosmic evolution for the long term is not well defined using dark energy in the consensus model, and is usually explained using anthropic arguments (the idea that, out of a multiverse with a horde of different properties, our universe is most likely to harbor life due to its own specific characteristics, and so the other universes are not observable) to prevent other possibilities from taking shape. The problem that this creates is that this hypothesis is not verifiable in any measurable way, and so is regarded as an unscientific explanation.

Furthermore, as mentioned previously, there is a multitude of possibilities that can occur in terms of the evolution of the cosmos, all of which can occur infinitely many times, and so the process of inflation can repeat itself. As a result, while its power of predicting a nearly scale-invariant spectrum of perturbations is well appreciated, it is not nearly as predictive about cosmic evolution. Therefore it is clear that a change is required in this regard.

## IV. The cyclic model presented in the context of the strengths and problems of inflationary theory

As explained in section II, the cyclic model developed by Turok and Steinhardt provides a new perspective to the old idea that the universe can get recycled periodically. However, this picture is based on an M-theory model – the universe that we observe around us is on one of two branes that move along a fifth dimension. The collapse of this interbrane separation to zero and the subsequent rebound is the event that is interpreted as a big crunch-big bang transition. While problems are clearly being faced in finding a detailed and specific understanding of the dynamics of the said transition, the ekpyrotic/cyclic picture does provide well-defined cosmic evolution over large time scales and dark energy, in contrast to the role it plays in the consensus model, is the key to ensuring that these cycles are self-sustaining.

At the observed 4 dimensions, dark energy is known to play the role of accelerating cosmic expansion, so that the branes become smooth, flat and nearly vacuous. This sets the model up for the contraction phase and the subsequent transition to expansion, which can be explained in the following way. As a quotidian example, consider a spring in the shape of a 'U'. Let this spring be twisted about itself and then released. It is observed that the closer the spring gets to its original shape, the faster it moves and it eventually attains equilibrium at that original shape, just as in the simple harmonic motion of a pendulum. Now let it be considered that the spring is the physical manifestation of dark energy in the cyclic model. It can be seen [3] that the interbrane distance increases and reduces almost exactly as in the analogy above. Certainly it could be argued that, due to gravity, the periodic expansion and contraction of this system could eventually stop, but this is prevented due to the conversion of some gravitational potential energy to brane kinetic energy, and this is made up for by the generation of matter and radiation during each cycle, a process that is interpreted in this theory as a big bang. It is therefore seen that dark energy plays a central role in the cyclic picture and is not seen as just another parameter added to make the theory commensurate with astronomical observations. Also, it is shown through the aforementioned analogy of the movement of the spring that the increase and reduction in the dominance of dark energy can be presented simply and intuitively, and something that is not just possible, but also physically realistic.

It has also been observationally proven that $V(\phi)$ is compatible with the requirements placed on the equation of state, $w_\phi$. At the current epoch, the value of $w(\phi)$ is placed at –0.8 [3], which is well within the limits previously placed for the acceleration of expansion $(-1 \leq w_\phi < -\frac{1}{3})$. Along with this, it is known that the scalar field $\phi$ determines the interbrane distance, so it can be seen that this field and the potential associated with it simultaneously solve both corresponding issues that were thought to be separate in the consensus model, namely the inflaton and dark energy. However, there is another issue concerned with the presence of $V(\phi)$ in the cyclic model – the true origin of this potential is not immediately known. Nevertheless, this once again poses a smaller problem than it would in the inflationary picture. If the brane worlds were created at a finite point in the past, then it would be assumed that the scalar field involved would also be created at that point, thereby increasing the energy requirement at that point, and without having any true mathematical proof that this is indeed the case. However, in this case, the problem is diluted, since the brane worlds have existed for an infinite period in the past, and so the issue of this potential being created at a finite time is removed.

The problem that plagued previous cyclic theories was the fact that, when the universe collapsed to a singularity, the entropy seemed to run down nearly to zero as time moved in the positive direction. Entropy evolution in this cyclic model is less of a problem than it is in previous ones or even in the consensus picture because, while dark energy is described here by $V(\phi)$, where $\phi$ acts along the extra dimension, the entropy itself is spread out across the two branes during and beyond the time of the big crunch-big bang transition. It also increases with each cycle as matter and radiation are generated at each bounce, and the branes keep expanding slowly through it.

As well defined as the cyclic picture might be, it still suffers from familiar problems – the dynamics of the bounce itself are not clearly defined, especially in quantum terms, and the potential $V(\phi)$ is yet to be derived quantitatively straight from established postulates. It is hoped that progress made in these areas will help to make the picture clearer, and help to present it as a more compelling candidate for a model of cosmology.

Despite the blemishes in this alternate model, there are two more points that make it a convincing contender as a description of the cosmos. The first deals with the postulated existence of the second brane as a sufficient way to deal with the complication of the singularity. If there were only one brane, an infinitesimal point of space-time would be unavoidable. With two branes, the question of infinite density, and hence infinite curvature, is removed. Gravity is also no longer acting at a single point, but instead is spread out, which also removes infinities from the dynamics of the interaction. For the question of why our universe is present on this particular brane and not the other, the anthropic principle is not required because there are only two possibilities of which the visible universe must be one. Furthermore, the second brane is not visualized because it is distanced along an extra dimension, which itself cannot be visualized or moved along.

Dark matter's presence in the universe has been explained through the significant role it plays in large-scale structure formation, but it has not been reasoned as an integral constituent of the evolution of the universe itself over larger time scales, either in inflationary theory or in the cyclic model. Whereas in the former, it is not absolutely required for inflation to work, and hence is simply added as another component (and has little consequence after the end of the matter-dominated epoch), a physical reason can be provided for its existence using the cyclic theory in the following way. During the quintessence-dominated phase, when dark energy spreads out all normal matter, dark matter plays a role in making the brane kinetic energy non-negligible and in making $V(\phi)$ decrease towards a negative value because it is posited that the energy density should be conserved in interactions between dark matter, the scalar field $\phi$ and dark energy (as described by $V(\phi)$). As the universe becomes more vacuous, dark matter also gets spread out more evenly and so, in order to maintain a certain level of energy density, energy is transferred to the kinetic energy of the field $\phi$, thereby starting the contraction phase. At the bounce, when gravitational potential energy is converted back into brane kinetic energy, some part of this is transferred back to dark matter so that all energies are reset to their original values at the start of each cycle. The reason dark matter is not confined in this picture to the extra dimension is twofold – firstly, the energy values arise directly from the dynamics of the interactions themselves and fine-tuning of these values by hand is avoided, therefore making this scenario as physically realistic as possible. Secondly, the existence of dark matter was found through astronomical observations in the normal 4 dimensions, which leads us to believe that it should have dynamical properties in these dimensions.

## V.  Conclusion and Implications

There are two major points to be drawn from this study. Firstly, after a review of inflationary cosmology as it stands today, a possibility has been considered that the dark energy observed today is a remnant of the primordial inflation energy, i.e. it is the same energy at a lower energy scale. The impact of inflation in predicting cosmic evolution over large time scales is weighed against that of the cyclic theory. The viability of the potential from each of the theories to be the constituent of the equation of state for dark energy is also evaluated. Secondly, I have attempted to provide qualitative reasoning to explain the mechanism of dark energy and the existence of dark matter using the cyclic model as a means of perspective. It is expected that the mechanism for the latter, which has been outlined in this paper, will be elaborated on in a future work.

The implications of this paper are thereby twofold. From the review of the consensus model, together with work to be done in the future, it is expected that inflation could be made a more interconnected picture, wherein the two separate sources of expansion in different epochs can hopefully be connected in some form. From the cyclic picture, it is projected that dark matter could be made an as integral part of the theory as dark energy, therefore giving physical meaning for each presently observed constituent of the universe.